\documentstyle{l-aa}

\begin{document}

   \thesaurus{12         
              (02.07.1;  
               12.04.1;  
               12.07.1)} 
   \title{The gravitational deflection of light in MOND}

   \author{B. Qin$^1$, ~X. P. Wu$^{1,2,3}$, ~Z. L. Zou$^{1}$
          }

   \offprints{X. P. Wu$^1$}

   \institute{$^1$Beijing Astronomical Observatory,
	     Chinese Academy of Sciences, Beijing 100080, China\\
              $^2$DAEC, Observatoire de Paris-Meudon,
              92195 Meudon Principal Cedex, France\\
              $^3$Department of Physics, Tanjing Normal University,
              Tanjing 300074, China
             }

  \date{Received ~~~~~~~~~~1994; accepted ~~~~~~~~~~~1994}

   \maketitle

   \begin{abstract}

   The deflection angle $\Delta\phi$ of light rays by the gravitational
   field of a spherical system $M(r)$ is calculated using the MOdified
   Newtonian Dynamics (MOND).
   It is shown that $\Delta\phi$ with an impact parameter $r_0$
   can be expressed by the measured rotation velocity $v(r)$ as
   $$
   \Delta\phi(r_0)=2\int_{r_0}^{\infty}\frac{v^2(r)}{c^2}
   \frac{r_0dr}{r\sqrt{r^2-r_0^2}},
   $$
   where
   $$
    v(r)=\left\{
   \begin{array}{ll}
   \left(Ga_0M(r)\right)^{1/4}, & ~~~~~r_0>r_c;\\
   \left(\frac{GM(r)}{r}\right)^{1/2},& ~~~~~r_0\leq r_c,
   \end{array} \right.
   $$
   and $r_c$ is the critical radius that is determined by the critical
   acceleration $a_0$.
   In the Newtonian limit of the gravitational acceleration $a\gg a_0$,
   $\Delta\phi$ approaches
   $\Delta\phi=2Gm(r_0)/c^2r_0$ with the projected surface mass $m(r_0)$.
   Whilst the asymptotic value of $\Delta\phi$
   reaches a constant $\pi(v_{\infty}/c)^2$ in the low-acceleration
   limit of $a\ll a_0$.
   Taking the empirical correction of a factor of 2  from the
   theory of general relativity into account and utilizing the relation
   between rotation velocity $v$ and velocity dispersion $\sigma$, MOND
   results naturally in a constant deflection angle,
   $4\pi(\sigma/c)^2$, which has been widely  used in the
   present-day study of gravitational
   lensing by galaxies and clusters of galaxies, implying that
   without introducing the massive halos acting as $r^{-2}$ for dark matter
   MOND has no difficulty in reproducing the known cases of gravitational
   lensing associated with galaxies and clusters of galaxies.

   \end{abstract}

%

\section{Introduction}

	Many attempts have been made for  decades to account for
the  large discrepancies
between the luminous matter seen from the galaxy radiations and the
dynamical matter derived from galaxy rotation curves.
A conventional way out of
these mass discrepancies is to assume the existence of some kinds of
unseen matter, which has been widely accepted today
by the community of astronomers and physicists and many experiments
have been devoted to searches for these dark matter despite that
there have been no any positive results obtained thus far.
Whilst some alternatives to the puzzle are to modify the Newtonian
inverse square law, among which the most successful of the empirically
suggested models is the Milgrom's MOdified Newtonian Dynamics (MOND)
(Milgrom 1983a, b, c).

	As a proposal to avoid introducing dark matter
in astrophysics, MOND has naturally predicted the flat form of the
rotation curves for spiral galaxies, the Tully-Fisher law and the
luminosity-rotation velocity/velocity dispersion relations in
spiral/elliptical galaxies.  Although there have been some arguments,
besides the unknown details of any fundamental physics,
against MOND using the stellar kinematics in dwarf galaxies
(Lake 1989; Gerhard \& Spergel 1992) and
the dynamical properties of X-ray selected clusters of galaxies
(Gerbal et al. 1992),  these claims still need to be further
considered (Milgrom 1991; 1993).

As was pointed out recently by Sanders and Begeman (1994), the more
fundamental problem with MOND is the observation of gravitationally
distorted images of background galaxies by the gravitational potential of
intervening clusters of galaxies.  Indeed, the deflection of light rays
by the gravitational field has remained to be an unsolved problem in
MOND (Milgrom 1993, private communication).
The presently known gravitational cases,
such as  the lensed quasar pairs, luminous arcs (see Surdej \& Soucail 1993 for
a summary), quasar-galaxy associations (see Wu 1994 for a summary), etc.,
associated with galaxies and clusters of galaxies
where MOND would appear to be important
can constitute a critical test for MOND.

This paper is then to calculate the deflection of light by the gravitational
field of a spherical matter system in MOND. Due to the absence of
a general theory of gravity in MOND comparable to General Relativity,
the present paper deals with the motion of particles only in its classical way.

%

\section{The equation of motion in MOND}

In MOND the true gravitational acceleration ${\bf g}$ is related to the
Newtonian gravitational acceleration ${\bf g_n}$ as
\begin{equation}
\mu(g/a_0){\bf g}={\bf g_n},
\end{equation}
where $a_0$ is a new fundamental constant or the critical acceleration
parameter with the value of $a_0\sim10^{-8}$ cm s$^{-2}$ (Milgrom \&
Braun 1988), and $\mu$ is some function that has the asymptotic behaviour
\begin{equation}
\begin{array}{ll}
\mu(x)=1, & \;\;\;\;\;\;\; x\gg1;\\
\mu(x)=x, & \;\;\;\;\;\;\; x\ll1.
\end{array}
\end{equation}
In particular, the major properties of the results in MOND are
insensitive to the choice of $\mu$ (Milgrom 1983b).  For simplicity,
we take $\mu(x)=1$ when $x>1$ and $\mu(x)=x$ when $x\leq1$.

The motion of a test particle  in the gravitational field
of spherical symmetry is described by
\begin{equation}\left\{
\begin{array}{lll}
a_r\mu(\frac{a}{a_0})& = & -\frac{GM(r)}{r^2},\\
& & \\
a_{\phi}\mu(\frac{a}{a_0}) & = & 0.
\end{array}\right.
\end{equation}
Here $a_r$ and $a_{\phi}$ represent the components of acceleration along
the $r$- and $\phi$- direction, respectively.  $M(r)$ is the total
mass enclosed within the radius of $r$. The equation of motion
of the test particle can then be specifically written as
\begin{equation}
\left\{
\begin{array}{ll}
\ddot{r}-r\dot{\theta}^2=-\sqrt{\frac{Ga_0M(r)}{r^2}}, & \\
                         & \;\;\;\;\;\;\;\;                   r> r_c\\
r^2\dot{\theta}=c_1;  &
\end{array}  \right.
\end{equation}
and
\begin{equation}
\left\{
\begin{array}{ll}
\ddot{r}-r\dot{\theta}^2=-\frac{GM(r)}{r^2}, & \\
                                & \;\;\;\;\;\;\;\;  r\leq r_c\\
r^2\dot{\theta}=c_2;  &
\end{array}  \right.
\end{equation}
where $r_c$ measures the radius separating Newtonian mechanics from MOND
and is determined by
\begin{equation}
r_c=\sqrt{\frac{GM(r_c)}{a_0}},
\end{equation}
$c_1$ and $c_2$ are the integral constants,  and
the time derivative of the position of the test particle
is denoted by ``$\cdot$".
The circular rotation velocity $v(r)$ resulting from these two different
kinds of equations of motion reflects their significant difference. Setting
$\dot{r}=\ddot{r}=0$ yields
\begin{equation}
v(r)=\left\{
\begin{array}{ll}
\left(Ga_0M(r)\right)^{1/4}, & \;\;\;\;\;\;\;\; r>r_c;\\
  & \\
\left(\frac{GM(r)}{r}\right)^{1/2}, & \;\;\;\;\;\;\;\; r\leq r_c.
\end{array} \right.
\end{equation}
For a pointlike mass or a finite mass distribution within a radius of $R$,
a constant circular velocity instead of the decreasing form of
$r^{-1/2}$ in the Newtonian mechanics is predicted by MOND when
$r$ tends to infinity. This
provides a natural explanation for the flatness of the rotation curves of
spiral galaxies alternative to dark matter.

Some initial and/or boundary conditions are needed to find
the solution to the orbit of a test particle from eqs.(4) and (5).
Define an impact parameter $r_0$
\begin{equation}
\dot{r}(r_0)=0
\end{equation}
so that the velocity of the test particle at $r=r_0$ can be written as
\begin{equation}
v_0=r_0\dot{\theta}(r_0)
\end{equation}
Moreover, the first-order time derivatives of position of the test particle
at the
critical radius $r_c$ are required to keep their continuities when the particle
crosses the boundary of $r=r_c$ from either sides.
An immediate consequence of this requirement
is that the two integral constants $c_1$ and $c_2$ in eqs. (4) and (5)
have the same value and
\begin{equation}
c_1=c_2=v_0 r_0
\end{equation}
In fact, this is the result of the conservation of angular momentum
analogue to the one in Newtonian dynamics.

The equation of motion of the test particle, eqs.(4) and (5), can
be unified to one simple equation
by replacing the matter distribution $M(r)$ by the
circular rotation velocity $v(r)$ of eq.(7)
\begin{equation}
\frac{d^2\left(\frac{1}{r}\right)}{d\phi^2}+\frac{1}{r}\left[
1-\left(\frac{r}{r_0}\right)^2\left(\frac{v(r)}{v_0}\right)^2\right]=0.
\end{equation}
The solution to this equation with the boundary condition of eq.(9) is
\begin{equation}
\frac{d\phi}{dr} = \pm \frac{r_0}{r\sqrt{r^2-r_0^2}}\frac{1}
{\sqrt{1-\frac{r^2}{r^2-r_0^2}\int_{r_0}^r\frac{2v(r)^2}{v_0^2}\frac{dr}{r}}}.
\end{equation}
Given the observed rotation curve $v(r)$ in and around the massive system or
the matter distribution $M(r)$, the orbit of
a test particle can be found from eq.(12).

%
%

\section{The deflection of light  in  MOND}

The deflection angle of light rays propagating from and to the distant
universe with an impact parameter of $r_0$ from the massive body $M$ is
\begin{equation}
\Delta \phi=2|\phi(r_0)-\phi(\infty)|-\pi.
\end{equation}
The test particle discussed in the above section is now replaced
by the photons and
hence, the speed of light $c$ is used instead of $v_0$.

Eq.(12) can now be expanded in the series of $(v(r)/c)^2$:
\begin{eqnarray*}
\phi(r)&= &\pm\left[\int_{r_0}^{r}\frac{r_0dr}
                    {r\sqrt{r^2-r_0^2}}\right.
\end{eqnarray*}
\begin{equation}
   \;\;\;\;\;\;\;\;\;\; \left.
  +\frac{r_0r}{(r^2-r_0^2)^{3/2}}\int_{r_0}^r\frac{v^2(r)}{c^2}\frac{dr}{r}
+O(\left(\frac{v(r)}{c}\right)^4)\right].
\end{equation}
Note that the integral of $\int_{r_0}^r\frac{v^2(r)}{c^2}\frac{dr}{r}$
approaches infinity when
$r\longrightarrow\infty$, which may then invalidate the expansion of
eq.(12) into eq.(14).  To see this, utilizing a constant
circular rotation velocity $v_{\infty}$
at a large distance ($r^*$) well beyond both
the radius of the massive body and the MOND limit leads to
$$
\int_{r^*}^r\left(\frac{v_{\infty}}{c}\right)^2\frac{dr}{r}
=\left(\frac{v_{\infty}}{c}\right)^2\ln\frac{r}{r^*}.
$$
However, the distance $r$ that satisfies $r\sim {r^*}
exp((c/v_{\infty})^2)$ would exceed the whole visible size of
the present universe for any realistic gravitationally bound systems like
galaxies and clusters of galaxies.
Therefore, eq.(14) is valid in the presently observable
universe.  The first term of the right-hand side of eq.(14) is the
orbit of a straight line when the photon travels in an Euclidean space,
and the rest terms represent the contributions of the gravitational field.

Finally, to the first order of $(v(r)/c)^2$ the deflection of light is
\begin{equation}
\Delta\phi=2\int_{r_0}^{\infty}\frac{v^2(r)}{c^2}
\frac{r_0}{r\sqrt{r^2-r_0^2}}dr.
\end{equation}

\subsection{Pointlike mass}

In the case of pointlike mass, inserting eq.(7) with
$M(r)=M$ into  eq.(15) yields:\\
(1)For $r_0>r_c$,
\begin{equation}
\Delta \phi=\pi \frac{\sqrt{Ga_0M}}{c^2};
\end{equation}
(2)For $r_0\leq r_c$,
\begin{eqnarray*}
\Delta\phi&= & \frac{2GM}{c^2r_0}\sqrt{\frac{r_c-r_0}{r_c+r_0}}
           +\frac{2GM}{c^2r_c}\sqrt{\frac{r_c-r_0}{r_c+r_0}}
\end{eqnarray*}
\begin{equation}
  \;\;\;\;\;\;\;\;\;\;\; +\frac{2\sqrt{Ga_0M}}{c^2}\sin^{-1}\frac{r_0}{r_c}.
\end{equation}
In particular, eq.(17) reads
\begin{equation}
\Delta\phi=\frac{2GM}{c^2r_0}, \;\;\;\;\;\;\; r_c\longrightarrow \infty,
\end{equation}
i.e.,the result of Newtonian limit. Whilst taking $r_c=r_0$ in eq.(17)
identifies the result in a non-Newtonian limit [eq.(16)].

The most interesting result appears in the low-acceleration limit or
at large radii of galaxies and clusters of galaxies where MOND plays
an important role. Eq.(16) can be simplified using the observable
asymptotic rotation velocity $v_{\infty}$
\begin{equation}
\Delta\phi=\pi\left(\frac{v_{\infty}}{c}\right)^2.
\end{equation}
The significant feature of the deflection of light in MOND is
its constant angle of light bending, in accord with the
flat rotation curves at large distances from the centers of galaxies
and of clusters of galaxies. A relatively
larger deflection  is then provided by MOND than that by the
Newtonian mechanics (see Fig.1).

%
   \begin{figure}
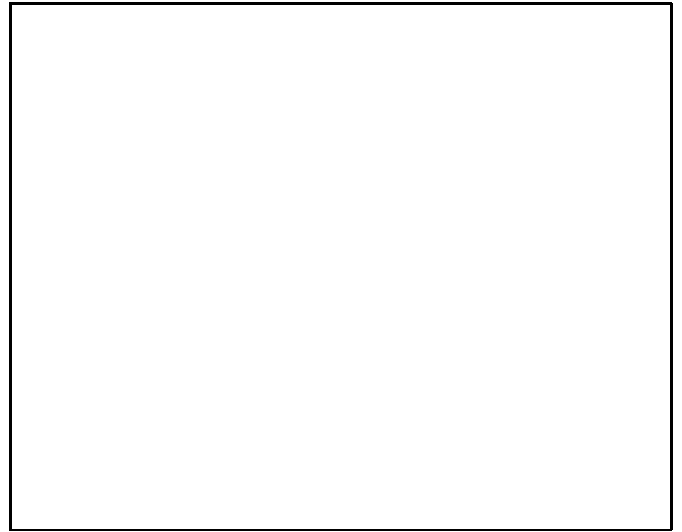

      \picplace{7cm}
      \caption{The deflecting angle of light $\Delta\phi$ in unit of
               $\phi_0=2\pi(v_{\infty}/c)^2$ by a pointlike mass versus
               the impact parameter $r_0$ in unit of $r_c$.  Beyond
               the critical distance $r_c$, MOND predicts
               a constant deflection of light instead of the declining
               form of $1/r$ derived from the Newtonian dynamics
               (the dotted line)
              }
         \label{Fig.1}
   \end{figure}

Recall that the present theory of gravitational
lensing uses indeed a constant deflection of light for the
dark matter associated with
galactic halos (Turner, Ostriker \& Gott, 1984) and for matter
distributions in clusters of galaxies (Wu \& Hammer, 1993; Wu 1993).
Essentially, these constant deflections of light can account for
the statistical properties of the lensed quasar pairs and the
giant luminous arcs.
To quantitatively compare the result of MOND with the update
theory of gravitational lensing that adopts the line-of-sight
velocity dispersion for galaxies and clusters of galaxies, the
circular velocity $v_{\infty}$ in eq.(19) should be written in
terms of the observed line-of-sight velocity dispersion.  Unfortunately,
there has not been set up an analog of the virial theorem in MOND
(Milgrom 1983c). Additionally, the present discussion is confined
to a pointlike mass rather than an extended system, although
the point mass can be regarded as a good approximation to a  galaxy seen
at a large distance. Note that there is no massive dark halo
at all surrounding the luminous galaxy according to MOND.
Basically, in a singular isothermal sphere the circular velocity
is related to the line-of-sight velocity dispersion by $v=\sqrt{2}\sigma$.
Therefore, the gravitational deflecting angle by a singular isothermal
sphere used in gravitational lensing today can be written as
\begin{equation}
\alpha=4\pi\frac{\sigma^2}{c^2}=2\pi\frac{v^2}{c^2}.
\end{equation}
Here both $v$ and $\sigma$ in a singular isothermal sphere
are independent of radius and thereby, $v=v_{\infty}$. This property
is the unique among the various matter distributions (Wu 1993).

Due to the absence of the general theory of gravitation in MOND,
a ``quasi-general relativity" correction of a factor of 2 to the
light bending derived from a classical way in MOND is
empirically adopted. As a consequence of this, eq.(19) reads
\begin{equation}
\Delta\phi=2\pi\frac{v^2_{\infty}}{c^2},
\end{equation}
which reconcile the presently used formula in the
study of gravitational lensing by larger systems (galaxies, clusters
of galaxies, etc.).  The key point is that MOND provides a deflection
of light identical to the one by a singular isothermal sphere
without introducing the massive halos having the density form of $r^{-2}$.
It then turns out that
all the presently known cases of gravitational lensing associated
with galaxies and clusters of galaxies can be explained by MOND
using the luminous matter alone that concentrated in the central
regions.

\subsection{Extended mass distribution: rotation velocity}

For a galaxy whose circular velocity has been well measured,
$\Delta \phi$ can be straightforward obtained using eq.(15).
A typical rotation
curve exhibits an increasing form in the central region and
tends to a constant $v_{\infty}$ beyond a characteristic radius of $r_h$.
Two types of light orbit should be specified:\\
(1)For $r_0>r_h$,
\begin{equation}
\Delta\phi=\pi\frac{v^2_{\infty}}{c^2};
\end{equation}
(2)For $r_0\leq r_h$,
\begin{equation}
\Delta\phi=2\frac{v_{\infty}^2}{c^2}\left[\sin^{-1}\frac{r_0}{r_h}
+\int_{r_0}^{r_h}\left(\frac{v^2(r)}{v^2_{\infty}}\right)
\frac{r_0dr}{r\sqrt{r^2-r_0^2}}\right].
\end {equation}
Figure 2 illustrates the variations of $\Delta\phi$ with the impact
parameters for three kinds of rotation curves.  Note that utilizing the
observed rotation curves in the calculation of $\Delta\phi$ is independent
of the assumed dynamics of whether it is the Newtonian one or the MOND.

%
   \begin{figure}
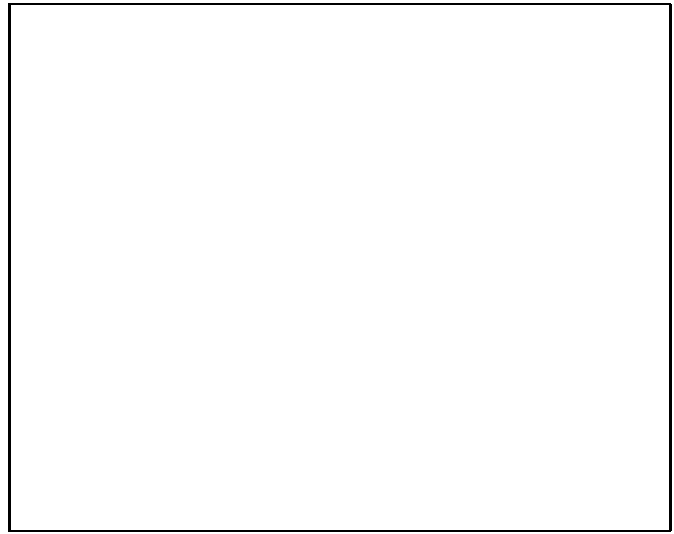

      \picplace{7cm}
      \caption{The deflecting angle of light $\Delta\phi$ in unit of
               $\phi_0=\pi(v_{\infty}/c)^2$ derived from the observed
               circular velocity. Three types of rotation
               curves in the central region of radius of $r_h$
               are assumed:
               (1)$v(r)=v_{\infty}(r/r_h)^{1/2}$ (top);
               (2)$v(r)=v_{\infty}(r/r_h)$ (middle);
               (3)$v(r)=v_{\infty}(r/r_h)^2$ (bottom).
               Beyond
               the radius of $r_h$ the rotation velocity remains
               unchanged, leading to a constant deflection of light.
              }
         \label{Fig.2}
   \end{figure}

\subsection{Extended mass distribution: surface brightness}

Surface brightness (optical/X-ray) of galaxies and of clusters of galaxies
are relatively easier to measure than their rotation velocities. Assuming
that the light profile traces the mass distribution as argued By MOND,
one can   obtain the deflection of light from eqs.(7) and (15):\\
(1)For $x_0\leq x_c$,
\begin{eqnarray*}
\frac{\Delta\phi(x_0)}{\phi_0}& = &\frac{x_cx_0}{\sqrt{\tilde{M}(x_c)}}
 \int_{x_0}^{x_c}\frac{\tilde{M}(x)dx}{x^2\sqrt{x^2-x_0^2}}
\end{eqnarray*}
\begin{equation}
 \;\;\;\;\;\;\;\;\;
 + x_0\int_{x_c}^{\infty}\frac{\sqrt{\tilde{M}(x)}dx}{x\sqrt{x^2-x_0^2}};
\end{equation}
(2)For $x_0>x_c$,
\begin{equation}
\frac{\Delta\phi(x_0)}{\phi_0}=
 x_0\int_{x_0}^{\infty}\frac{\sqrt{\tilde{M}(x)}dx}{x\sqrt{x^2-x_0^2}}.
\end{equation}
Here $\phi_0=4\sqrt{Ga_0M_T}/c^2$, and $\tilde{M}(x)$ is the
dimensionless mass enclosed within $x$
normalized to the total mass $M_T$. All the
distances ($x$) are measured in unit of the so-called length scale
in the model fit to the surface luminosity distribution. The subscripts
``$_c$" and ``$_0$" indicate the critical  and the impact
positions, respectively.

An example is shown in Fig.3 using the King model for the matter
distribution.
This model  has been found to fit fairly well the surface
brightness of the central parts of galaxies and of clusters of galaxies,
and has been widely adopted today in the modeling of the luminosity
distribution for these systems.
The spatial mass distribution $M(r)$ can be found to be
$$
M(x)=\frac{9b\sigma^2}{G}\left(\ln(x+\sqrt{1+x^2})
-\frac{x}{\sqrt{1+x^2}}\right),
$$
in which $x=r/b$ and $b$ is the core radius.
The deflection of
light resulting from this model shows a slight increasing
form instead of the declining one of $\sim r^{-1}$ in the Newtonian
limit, when the impact distance tends to infinity. This shows again that
MOND provides a significantly large deflection of light beyond $a_0$..

%
   \begin{figure}
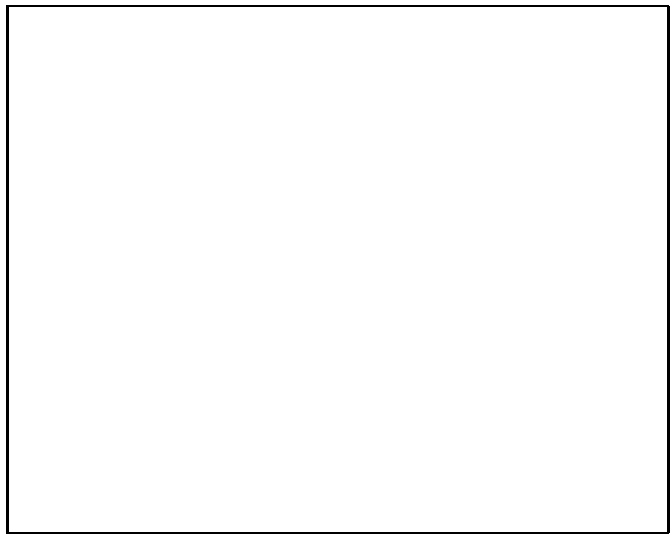

      \picplace{7cm}
      \caption{The deflection of light $\Delta\phi/\phi_0$
	       by the matter distribution following a King model.
               Four critical radii in unit of core radius
               $b$ have been shown: (from top to bottom) $r_c=100b$,
               $r_c=10b$, $r_c=5b$ and $r_c=b$.
               The normalization $\phi_0$ is
               $(9a_0b\sigma^2)^{1/2}/c^2$. The results from
               the Newtonian dynamics are also plotted in the dashed
               lines for comparisons.
               }
         \label{Fig.3}
   \end{figure}

%
%

\section{Discussion and conclusions}

Dark massive halos of galaxies and unseen matter excess in clusters of
galaxies, if they were real,
can be well described by a singular isothermal sphere
with a power law of $r^{-2}$. This model has been widely used in
the study of gravitational lensing today. Certainly, this is
due partially to the fact that  it results
in a constant deflection of light $4\pi(\sigma/c)^2$, making the
calculations much simplified.  Nevertheless, there have been no
any conclusive  evidences, from the experiments that search for dark
matter, for the existence of the hidden mass.  Dynamical analysis
of the rotational curves of galaxies on the basis of Newtonian
mechanics is the only strong observational
evidence that shows the mass discrepancies.

An alternative to the missing mass is to modify the Newtonian gravity.
Though it is not conventional,
Milgrom's MOND has successfully predicted the flat rotation curves
of galaxies, leading to a new sight into the dynamics on large
scales. The present paper has computed one of the critical issues
in MOND, the deflection of light rays by a spherical gravitational field.
Surprisingly, MOND provides a constant deflecting angle at large
distance from the center of the gravitational field, which
is consistent with the
the value presently used in the study of gravitational lensing by galaxies
and  clusters of galaxies. It is then likely
that all the lensing cases can be equally reproduced in MOND without
the massive dark matter in galaxies and in clusters of galaxies.
Similar to the flatness of rotation velocity in galaxies predicted
by MOND without assuming the
massive halos, the constant deflection of light from MOND
has the same effect as the $r^{-2}$ halos did.

It has been shown that light bending is no more a critical argument
against MOND today. Conversely, MOND predicts a reasonable deflection
angle of light by large massive systems. Therefore,
whether or not MOND reflects the nature of gravity needs
to be further investigated using other astronomical methods.

\begin{acknowledgements}
We thank Moti Milgrom for encouragement and comments.
This work was supported in part by China Natural Science Foundation.
\end{acknowledgements}

\end{document}